\pdfoutput=1

\documentclass[english]{lni}
\usepackage[utf8]{inputenc}
\usepackage[T1]{fontenc}
\usepackage{paralist}
\usepackage{xspace}
\usepackage{fancyvrb}
\usepackage{graphicx}
\usepackage[caption=false]{subfig}
\usepackage{booktabs}
\usepackage{placeins}

\usepackage{tikz}
\usetikzlibrary{arrows.meta}
\usetikzlibrary{positioning,fit}
\usetikzlibrary{backgrounds}
\usetikzlibrary{decorations.pathmorphing}
\usetikzlibrary{patterns}
\usetikzlibrary{shapes}
\usepackage[most]{tcolorbox}

\usepackage{etoolbox,xspace}
\usepackage[textsize=tiny,textwidth=3cm]{todonotes}
\usepackage[
  backend=biber,
  style=LNI
]{biblatex}
\usepackage{csquotes}
\MakeOuterQuote"

\newcommand{\cmd}[1]{\texttt{#1}}

\newcommand{\etal}[1]{et \emph{al.}\xspace}
\newcommand{\eg}[1]{e.g.\xspace}
\let\savesubsubsection\subsubsection
\renewcommand{\subsubsection}[1]{\savesubsubsection{#1}\vspace*{-1em}}
\let\savesubsection\subsection
\renewcommand{\subsection}[1]{\savesubsection{#1}\vspace*{-1em}}
\let\savesection\section
\renewcommand{\section}[1]{\savesection{#1}\vspace*{-1em}}

\graphicspath{{plots-tikz/}}
\let\pgfimage=\includegraphics 

\makeatletter
\newcommand\querysize{\@setfontsize\querysize\@vipt\@viipt}
\makeatother

\lstdefinestyle{query}{
  language=SQL,
  stepnumber=1,
  numbersep=10pt,
  tabsize=4,
  showspaces=false,
  showstringspaces=false,
  basicstyle=\linespread{1}\fontfamily{lmtt}\selectfont\querysize,
  keywordstyle=\color{blue},
  stringstyle=\color{purple},
  upquote=true,
  breaklines=true,
  commentstyle=\color{CadetBlue}
}

\definecolor{mygray}{rgb}{0.643,0.643,0.643}
\newtcolorbox{querybox}[2][]{%
  sidebyside align=top,
  enhanced,
  boxsep=0pt,
  arc=0pt,
  top=-3pt, bottom=-3pt,
  left=2pt, right=0pt,
  colback=white,
  colframe=mygray,
  boxrule=0.5pt,
  leftrule=12pt,
  overlay unbroken and first ={%
    \node[rotate=90,
          minimum width=0.5cm,
          anchor=south,
          font=\small\rmfamily,
          yshift=-13pt,
          white]
    at (frame.west) {#2};
  }
}

\begin{document}

%%% Mehrere Autoren werden durch \and voneinander getrennt.
%%% Die Fußnote enthält die Adresse sowie eine E-Mail-Adresse.
%%% Das optionale Argument (sofern angegeben) wird für die Kopfzeile verwendet.
%\title[Ein Kurztitel]{Tune out the Noise: Towards Deterministic Latencies in Query Processing for In-Memory Databases\protect\footnotemark}

%\title{Experimental evaluation of deterministic query latencies \break for in-memory database systems}
% title inspired by
% https://dl.gi.de/handle/20.500.12116/17321

%\title{Kondo your DB/OS System Software Stack}
%\subtitle{If it does not spark joy, discard.}
%\title{System Software Stacks for Data Processing \break in Multi-Tenant Environments with Real-Time Demands}

%Silentium! A step-wise Run--Analyze--Fix  to 
%Experimental Evaluation of Software System Stacks for blablabla

%\title{Slimming down the DB/OS Stack}
\title[Silentium! Run--Analyse--Eradicate the Noise out of the DB/OS Stack]{Silentium! Run--Analyse--Eradicate the Noise \break  out of the DB/OS Stack}

%%%\subtitle{Untertitel / Subtitle} % if needed
\author[W.~Mauerer   \and R.~Ramsauer  \and E.~Lucas \and D.~Lohmann \and S.~Scherzinger]
{Wolfgang Mauerer\footnote{Ostbayerische Technische Hochschule Regensburg, Germany \email{wolfgang.mauerer@othr.de}}\footnote{Siemens AG, Corporate Research and Technology, Munich, Germany} \and 
Ralf Ramsauer\footnote{Ostbayerische Technische Hochschule Regensburg, Germany \email{ralf.ramsauer@othr.de}} \and
Edson R. Lucas F.\footnote{Universit\"at Passau, Germany \email{edson.lucas@uni-passau.de}} \and 
Daniel Lohmann\footnote{Leibnitz Universit\"at Hannover, Germany \email{lohmann@sra.uni-hannover.de}} \and \break
Stefanie Scherzinger\footnote{Universit\"at Passau,  Germany \email{stefanie.scherzinger@uni-passau.de}} }

%%%%%%%%%%%%%%%%%%%%%%%%%%
% REVIEWS
%\pagenumbering{gobble}
%\include{response_reviewers}
%\pagenumbering{arabic}
%\newpage
%\setcounter{section}{0}
%\setcounter{page}{1}
%%%%%%%%%%%%%%%%%%%%%%%%%%

\startpage{1} % Beginn der Seitenzählung für diesen Beitrag / Start page
\editor{} % Names of Editors
\booktitle{} % Name of book title
\year{2021}
%%%\lnidoi{18.18420/provided-by-editor-02} % if known
\maketitle

\begin{abstract}
When multiple tenants compete for resources, database performance tends to suffer. 
Yet there are scenarios where guaranteed sub-millisecond latencies are crucial, such as in real-time data processing, IoT devices, or when operating in safety-critical environments.
In this paper, we study how to make query latencies deterministic in the face of noise (whether caused by other tenants or unrelated operating system tasks).
We perform controlled experiments with an in-memory database engine in a multi-tenant setting, where we successively eradicate noisy interference from within the system software stack, to the point where the engine runs close to \emph{bare-metal} on the underlying hardware. 

We show that we can achieve query latencies comparable to the database engine running as the sole tenant, but without noticeably impacting the workload of competing tenants.
We discuss these results in the context of ongoing efforts to build custom operating systems for database workloads, and point out that for \emph{certain} use cases, the margin for improvement is rather narrow.
In fact, for scenarios like ours, existing operating systems might just be \emph{good enough}, provided that they are expertly configured.
We then critically discuss these findings in the light of a broader family of database systems (\eg, including disk-based), and how to
extend the approach of this paper accordingly.
\end{abstract}

\newcommand{\query}[1]{\texttt{#1}}

\begin{keywords}
%HW/SW co-design \and
Low-latency databases \and tail latency \and real-time databases \and
bounded-time query processing \and DB-OS co-engineering
\end{keywords}

\section{Introduction}
% We could somewhere integrate a motivation for our work
% by paraphrasing the following by Goetz Graefe:

%And my standard analogy for that is if you own a house, every year you pay good money yet you hope you will never get anything back for it. It’s called fire insurance, right? You pay it, and you pay it willingly as long as it’s a small fraction of the value of the house. And as long as you can count on if you lose the house due to fire, it’ll get replaced. So, then similarly with robust performance, you’re probably willing to forego some small amount of efficiency, if, in return, you get robust performance, predictable performance, reliable performance, which among other things, permits you to load your service much higher. 

%If you get random load spikes and you never know when, you end up running a service at 20 percent utilization. But if your performance is very steady and utilization is very steady, it’s perfectly reasonable to run at 60 percent utilization. And suddenly, the ten percent overhead – or maybe even the factor two overhead – comes back and more, if you can load your service substantially higher.

The operating system is frequently considered boon and bane for the development of scalable service stacks.
While general-purpose operating systems (like Linux) provide a great deal of hardware support, drivers and system abstractions, they have also been identified as a cause of jitter in network bandwidth, disk I/O, or CPU~\cite{DBLP:journals/pvldb/SchadDQ10,armbrust09abovetheclouds,180314} when operating software services in cloud environments, where multiple tenants compete for resources. 
Naturally, this also affects the performance of cloud-hosted database engines~\cite{10.1007/978-3-319-15350-6_11}.
%However, in operating database engines, there are important use cases where  high throughput is not the sole optimization criterion, but where latencies must be low and deterministic, e.g., with real-time demands. 

Unacceptable noise and long-tailed latency distributions, but also the recent advances in hardware technology, have renewed interest in building database-specific operating systems.
While historically, database and operating-systems research have been highly interwoven, the communities have parted ways in the past, and are just now rediscovering potential synergy effects (e.g.~\cite{DBLP:journals/corr/abs-2007-11112,Muehlig2020}).
This has sparked immense interest in devising novel system architectures~\cite{mci/Kiefer2013}, especially for database-centric operating systems kernels (e.g.,~\cite{DBLP:journals/corr/abs-2007-11112,mueller:2019:sfma,Muehlig2020,DBLP:journals/debu/Giceva19,DBLP:conf/cidr/GicevaSSAR13}) that aim at deterministic performance.  
However, implementing an OS kernel is a herculean effort with tremendous follow-up costs, requiring substantial and largely duplicate effort for otherwise generic tasks, such as writing and maintaining device drivers, file systems, and infrastructure code, among others.

\vspace*{-1em}\paragraph{About This Paper.}
We %suggest to 
take a fresh look at standard operating systems for low-latency/high determinism workloads, as they arise in real-time scenarios. Similar  problems arise in cloud settings, where latency effects along the data path 
add up and can lead to substantial systemic problems, as Dean and Barroso have pointed out~\cite{10.1145/2408776.2408794}.
Rather than designing a new kernel from scratch%
\footnote{Whether to build a new operating system from scratch or whether to extend existing systems to cater to data processing needs has been an ongoing debate for decades~\cite{Gray1978}.}
to avoid noise and jitter, we follow an orthogonal % (yet complementary) 
approach, employing existing open-source components:
Identify the root causes, analyse, and then address them as far as possible within the \emph{existing} components. If necessary, enhance.

By vertical, cross-cutting engineering, we tailor the stack towards the needs of database
engines, eradicate interference, and ultimately, reduce any noise-induced latencies in query evaluation.  Our first results show that in many cases, a large degree of jitter is %already 
avoidable by the well-considered and purposeful employment of existing architectural measures -- actually measures originally developed for other domains, such as embedded real-time systems.
We present controlled experiments with an in-memory database engine running in a multi-tenant scenario on a number of different system software stack scenarios.
%, which are based on existing open source components.

We focus on in-memory database engines as a specific (and deliberately narrow) use case, as they are often employed in domains for which deterministic latencies are essential~\cite{DBLP:journals/arc/BuchmannL01}, and thus considered a particularly convincing use case for developing specific operating-systems or even a bare-metal database stack~\cite{254358,DBLP:journals/corr/abs-2007-11112,DBLP:conf/cidr/GicevaSSAR13}.
In this realm, our experimental setup, which is available as a Docker image for easy reproduction, can also serve as a baseline for researchers building special-purpose operating systems to compare their results against.
In particular, we claim the following contributions:

%NOT what we do, but what we show.

\begin{compactitem}
\item We perform controlled experiments with an in-memory database engine running on custom system software stacks based on existing open source components.
By careful cross-cutting engineering, we modify this stack to eradicate interference, and to ultimately reduce any noise-induced latencies in query evaluation. 

%Ultimately, we have the database engine running close to bare-metal on a dedicated CPU core, achieving deterministic latencies despite competing tenants executing on the same hardware. 

\item
We show that 
we can achieve the same performance using %customised
available operating systems as compared to running the database workload (near) \emph{bare-metal}.

\item
We show that
we can achieve the same performance in a multi-tenant scenario as compared to a database engine executing as the sole tenant without competing load.

\item 
We voice doubts whether these specific scenarios can benefit from operating systems custom-designed towards database workloads, as they are currently being proposed. 

%\item We establish that the necessary adaptions require advanced skills across the complete DB/OS stack. This overall non-trivial effort is rewarded by substantial improvements towards deterministic latencies.

%Considering multiple tenant scenarios sets this work apart from many typical benchmarks conducted in database research, where an engine is commonly benchmarked in isolation, which is not realistic for many application scenarios.

\item 
We discuss the potential generalisability of our approach to disk-based database engines, and systems involving I/O. In particular, we discuss opportunities that call for the joint efforts of the operating systems and database communities.
\end{compactitem}

\vspace*{-1em}\paragraph{Structure.} 
\looseness=-1 Our paper is structured as follows. We give an overview in Section~\ref{sec:prelims},
survey related work in Section~\ref{sec:related}, and
present our experiments in Section~\ref{sec:experiments}.  Their consequences are discussed in a more general context in Section~\ref{sec:discussion}. We conclude in Section~\ref{sec:conclusion}.

\section{Overview}
\label{sec:prelims}
We start with a brief
summary of possible perturbations of an executing database workload by neighbourly noise, followed by an overview of the system software stack scenarios
considered in this paper.
In this section and beyond, by the term \emph{kernel} we refer to the operating systems kernel (not the database kernel).

\subsection{Sources of Noise}
The three major sources of noise as observed by an unprivileged userspace
workload (as compared to system services or the kernel) are (1) other processes and
system services that compete
for CPU usage, (2) CPU performance optimisations (caches, pipelines, \dots)
that can usually not be disabled or controlled, and (3) contention of
implicitly shared resources (memory bus etc.). The signature of such
\emph{systemic} noise is not necessarily distinguishable from the
\emph{intrinsic} noise of the application, that is, variations in run-time
caused by data-dependent code paths, application-specific optimisations,
and so forth.

\vspace*{-1em}\paragraph{Processes and system services.}
Multi-tasking operating systems manage \(M\)
schedulable entities that compete for \(N\) processors,
with \(M \gg N\). Linux uses a \emph{completely fair scheduling}
(CFS)~\cite{mauerer2010professional} policy for regular processes, but also includes support for (soft) real-time scheduling via FIFO and round robin. The kernel can preempt
most userland activities (depending on the preemption model  statically configured at kernel build time), for instance upon the arrival of interrupts. It
can also place \emph{kernel threads} into the schedule that perform activities on behalf of
the kernel (for instance, to support migrating processes across CPUs, to perform post-interrupt
actions, etc.), and enjoy higher priority than regular processes,
regardless whether these are governed by real-time policies. 
The %influence and 
interplay of these factors creates noise compared to
an uninterrupted, continuous flow of execution of a single job.

\vspace*{-1em}\paragraph{CPU noise.} Even given the uninterrupted execution of code on a CPU,
pipelined and superscalar execution of code may lead to different temporal behaviour
than would be achieved by a straightforward execution of assembly instructions, which
manifests itself as another source of noise. Also, caching mechanisms (most importantly,
the cache hierarchy that comes into play with memory references, but possibly also
mechanisms like translation lookaside buffers used in virtual-to-physical address
translation) cause (widely) varying latencies in accessing memory.
This effectively
adds noise.

\vspace*{-1em}\paragraph{Shared resources.} Workloads executing on different CPUs are not entirely
isolated from each other, but interact via shared resources (cache, memory, etc.) that
are accessed via system buses. This even holds despite a possible
logical partitioning of system components that we discuss later. While the overall
situation (for instance, handling competing requests for bus usage) is deterministic
from a system-global view, delays caused by competing requests manifest as noise when viewed from the perspective of an individual process.
%%%%%%%%%%%%%%%%%%%%%%%%%%%%%%%%%%

\subsection{Experimental System Configurations}\label{sec:exp_conditions}
\begin{figure}[t]
\centering
\input{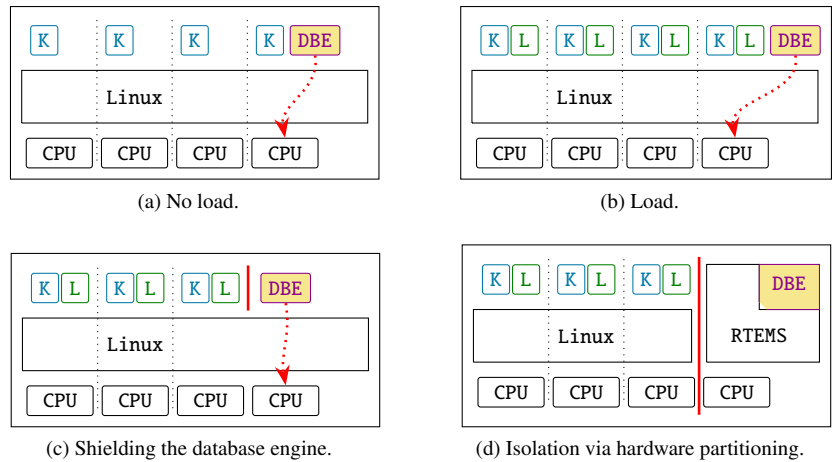}
\caption{System software stack scenarios.}
\label{fig:software-stack}
\end{figure}

The configurations of the system software stack, as used
in our experiments, are visualised in Figure~\ref{fig:software-stack}.
For now, we treat the in-memory database engine~(DBE) as a black box. %(Section~\ref{sec:experiments} provides details).

\smallskip
\noindent
\textbf{No Load.}
%\subparagraph{No load.}
In the \emph{no-load} scenario (Figure~\ref{fig:noload}), a single database engine executes on an otherwise quiet multicore system. The database payload is pinned to one CPU (c.f.\ the dashed arrow), to avoid perturbations, for instance caused by the scheduler moving the process across CPUs. However, standard system services, as limited to the bare necessity, and kernel threads required by the operating system proper (``K'' in the figure) can execute on all CPUs, including the CPU dedicated to the database workload.

\smallskip
\noindent
\textbf{Load.}
%\subparagraph{Load.}
In the \emph{load} scenario (Figure~\ref{fig:load}), additional tenants put the system under maximum strain. We simulate this payload (marked ``L'') by running synthetic workloads 
%(generated by  \verb!stress-ng!, as explained later) 
on each CPU.

While the database workload is again pinned to one particular CPU, it is scheduled by the operating system alongside kernel threads and the described payload. In the load scenario, we assume the viewpoint of a cloud provider maximising the utilisation of the available resources, while serving all tenants equally. We therefore refrain from assigning the database workload higher priority than the payload generated by the competing tenants.

\smallskip
\noindent
\textbf{Load/FIFO.}
%\subparagraph{Load/FIFO.}
A variation of the \emph{load} scenario uses standard Linux mechanisms to set a real-time scheduling policy for the database workload.
All load processes fall under scheduling policy ``other'', and compete for CPU resources as managed by the Linux standard scheduler. We place the database task in the real-time scheduling group SCHED\_FIFO, so it can preempt any other userland tasks that execute on the CPU. However, the database task can still be preempted by the kernel, or by incoming interrupts.

\smallskip
\noindent
\textbf{Shielding.}
%\subparagraph{Shielding.}
Another approach towards isolating the database workload from noise is to use CPU shielding (Figure~\ref{fig:shield}). 
This distributes all existing tasks and kernel threads on a given CPU to the rest of the system, and prevents utilisation of the shielded CPU by the standard scheduling for processes that are not explicitly assigned to this CPU.

We additionally make sure that incoming external interrupts only arrive at other CPUs.
Nevertheless, main memory, buses, caches, etc.\ remain shared resources in the system,  and accesses can induce additional noise that goes beyond the pure CPU noise. Additionally, the kernel can still preempt the
single running userland task (for instance, when timers expire), and latencies can arise from 
administrative duties performed by the kernel on such occasions, or in the context of system calls
issued by the task.

One set of measurements combines shielding and real-time priorities. This limits the kernel's abilities
to preempt the running userspace task. However, some caution needs to be administered: Not only
the ability to preempt a running task, but also the amount of work performed in kernel context when
a preemption occurs influences latency variance, and this amount is highly dependent on specific
(static) kernel configuration settings.\footnote{Linux provides a tick-less mode, which eliminates periodic interventions by a regular timer (at frequencies
ranging from 100Hz to 1,000Hz, depending on compile-time settings), but which may cause overhead
on other occasions, because maintenance of data structures performed during such ticks
must be performed ``en block''.}
Isolation in this scenario is based on guarantees provided by the Linux kernel. This implies trust in a complex, monolithic code base, which is undesirable for safety-critical scenarios.

\smallskip
\noindent
\textbf{Partitioning.}
%\subparagraph{Partitioning.}
The strongest form of isolation that we consider in this paper (Figure~\ref{fig:jailhouse}) relies on the Jailhouse hypervisor~\cite{DBLP:journals/corr/RamsauerKLM17}.
Jailhouse can partition system hardware resources by establishing independent and strictly isolated computing domains. 
Jailhouse leverages extensions of the underlying system architecture which include essential virtualisation mechanisms for system partitioning, such as segregation of CPUs, memory and devices, as well as additional extensions that allow to control the utilisation of shared resources, such as caches or system buses.

Jailhouse comes at a negligible performance overhead, as it does neither (para-)virtualise or emulate resources, nor schedule its partitions (\emph{guests}) among CPUs. The virtual machine monitor only interferes in case of critical exceptions and access violations.
This architecture can find application in multi-tenant database scenarios, described in~\cite{DBLP:conf/edbt/MuheKN12}, and in particular, safety-critical scenarios, which require %a strong
spatial-temporal isolation between tenants.

%Since DBToaster is a serverless database engine, with no dependencies beyond a C++/STL runtime environment, there is no need to deploy it on top of a full-blown general-purpose operating system such as Linux.
\smallskip
\noindent
\textbf{Bare-Metal Operation.}
Data center, cloud and high performance data processing systems often employ x86 server class
CPUs, and we have argued before that such use cases benefit from bounded tail latencies.
Other important use cases that require determinism are found in embedded systems, which
are typically equipped with ARM CPUs. Consequently, our investigation addresses both, x86 and ARM.

Using a simplistic ARM core that is just capable enough for realistic database deployment
reduces systemic noise that stems from multicore effects,
as found on server-class x86 CPUs~\cite{10.5555/77493}
to the bare minimum, (\eg, long pipelines, large caches,
and strong interference on buses). This allows us to explore
the \emph{intrinsic} variations of our database workload.

We employ the in-memory database engine DBToaster~\cite{DBLP:journals/vldb/KochAKNNLS14} (see also Section~\ref{sec:experiments}),  a highly portable serverless database engine that requires a C++/STL run-time environment, but no other libraries
or system services. Plain C++ can be executed without relying on an OS proper with moderate effort, but  the STL requires (at least conceptual) support for threads and preemptive locking, as well as
a full memory allocator. These requirements do not create a need to deploy it on top of a
fully-fledged general-propose operating system, such as Linux or Windows, but we deem
the implementation efforts large enough to warrant a tiny operating system.
Thus, we ported the database engine to RTEMS (real-time executive 
for multiprocessor systems)~\cite{10.1145/2597457.2597459}, a mature, tailorable embedded real-time operating system (with a 25-year development history) that finds deployment in
systems ranging from IoT devices to Mars orbiters.
Similar to unikernel approaches~\cite{madhavapeddy2013unikernels, bratterud2015includeos}, 
RTEMS and the database engine
are linked together into one single executable. This binary can be either booted as stand-alone
operating system on a bare-metal system, or (given low-level changes like the use of
a custom bootloader and adaptations of the RTEMS kernel to Jailhouse) be executed in 
parallel to Linux on a partitioned system
(as visualised in Figure~\ref{fig:jailhouse}).

To reduce operating system noise as far as possible, we essentially limit RTEMS to providing 
only a console driver,
and execute the database engine in a single thread, which eliminates the need for a scheduler.
This configuration is supposed to reduce any OS noise to the bare minimum, and is \emph{comparable
to a bare-metal},\unskip\kern-2pt\footnote{Bare-metal operation refers to code that runs without distinction between
payload and OS close to the hardware, without intermediary layers. This in contrast to,
for example, Ref.~\cite{10.1145/3209950.3209953}, which uses the term to denote code that 
runs without containers or virtual machines, but still relies on heavyweight, multi-million-LoC
OS kernels.} main-loop style binary.

%The database engine evaluated by us does not just rely on basic C++ runtime provisions, but %also
%makes heavy use of the C++ Standard Template Library (STL).

\section{Related Work}
\label{sec:related}

In this paper, we focus on in-memory database engines. 
We refer to~\cite{10.1109/69.180602} for an early overview of their architecture, and to~\cite{DBS-058} for a more recent survey.

In real-time scenarios, where in-memory database engines traditionally play an important role~\cite{DBLP:journals/arc/BuchmannL01}, deterministic latencies are crucial.
However, aspects such as consistency of query answers given transactional workload, or alternative tuple consumption strategies, are of no concern to the work presented in our paper, since we treat the database largely as a black box, and are interested in the overall system software stack. 

What is indeed highly relevant for us is the existing work on worst-case execution time (WCET) of queries in in-memory databases~\cite{DBLP:books/idea/encyclopediaDB2005/Buchmann05}, which considers control-flow graphs through the code (in fact, in the presentation of our experiments, effects of different paths through control flow graphs during query processing actually become visible).

Also close to our work in both methodology and context is research on the
influence of NUMA effects, focusing on in-memory database engines in particular. It is known that assigning threads to CPUs improves database performance, due to caching effects~\cite{mci/Kiefer2013,10.1007/978-3-319-15350-6_11,8509271}. Similar studies of assigning database workloads to computational units 
can be found throughout database research, for instance in
Refs.~\cite{DBLP:journals/pvldb/PorobicPBTA12,10.1145/3076113.3076121}.
Our experiments also assign threads to dedicated CPUs, and we benefit from data caching, but our motivation differs, as we  \emph{isolate} the database workload from harmful noise.
%generated by competing workload.

Databases operating in multi-tenant environments are another focus of our work. This differs from many benchmarks conducted in database research, where database workload often runs in isolation, while multi-tenant environments are closer to real-world conditions.
Similarly, an overview over \emph{performance isolation} for cloud databases is provided in~\cite{10.1007/978-3-319-15350-6_11}.

A systematic discussion of multi-tenant in-memory databases is provided  in~\cite{DBLP:conf/edbt/MuheKN12}:
From the viewpoint of a cloud provider, guaranteeing narrow service-level-agreements is a challenge, since
the provider must cater to all tenants, while utilising the  hardware resources. This mindset is also found in engineering for mixed-criticality systems~\cite{vestal:07:rtss,burns:18:acmcs}, where a critical workload (in our case, the database engine) must be shielded from noise (in our case, competing tenants), without cutting into the performance of the remaining workload.

In designing multi-tenant database engines, shielding tenants can be implemented on several levels in the system software stack.
Aulbach~\etal~\cite{DBLP:conf/sigmod/AulbachGJKR08} enable multi-tenancy on the level of the database schema; by appropriately mapping between the tenants' schemas and the internal schema, tables may be transparently shared between tenants. By rewriting queries, the authors ascertain isolation between tenants in an otherwise standard database engine.

Narasayya \etal~\cite{narasayya2013sqlvm} also aim at resource isolation, for the database-a-service provider Microsoft SQL Azure. They explore virtual machine mechanisms in userland without
relying on mechanisms provided by the kernel (and, consequently, not benefiting
from the guarantees provided by the OS kernel -- for instance, some isolations are not possible in userland, such as access to shared buses and other resources). 

Noll \etal~\cite{8509268} discuss how to accelerate 
concurrent workloads inside a single database engine by partitioning caches.
This feature is not targeted at multi-tenant databases per se, but applicable in general. However, this feature is specific to Intel CPUs. Further, it  is not directly subject to control
from userland, but exposed to applications by the
\verb!sysfs! pseudo-filesystem interface of the Linux kernel. Our x86-based RTEMS measurements in
a Jailhouse cell actually use the same infrastructure to assign a portion of the
cache to the system performing the measurements, which
reduces variations in memory access times. 

The general idea of using existing OS-level isolation mechanisms to reduce the amount of inference between latency (or otherwise) sensitive database workloads and the rest of a system has also been
pursued by Rehmann \etal~\cite{10.1145/3209950.3209953}: The authors use Docker containers to
isolate database instances from system and competing payload noise. 
Their work essentially implements limiting the CPU quota available to tasks, and pinning database-relevant operations to specific CPUs in the system. 
Especially the latter is similar to some of our experiments, albeit we
additionally include scheduling prioritisation and control the system noise
on pinned CPUs with various measures.
Thus, we make use of a richer toolset to achieve stronger levels of
isolation, as our measurements show. In fact, containers are conceptually
not intended to isolate a given workload from other workloads, but to provide a
specific, probably restricted  view of the system to a given workload.

Currently, there is renewed interest in building database-specific operating systems, partly motivated by such problems as unpredictability in performance.
For instance, the MXKernel project~\cite{Muehlig2020} proposes an alternative to the classic thread model, to cater to the demands of large-scale data processing. The
 DBOS initiative~\cite{DBLP:journals/corr/abs-2007-11112} goes so far as to envision managing database-internal data structures inside the OS kernel.
 Further, there are suggestions to share the database cost model with the operating system~\cite{DBLP:conf/cidr/GicevaSSAR13}, to allow for more transparency and to ultimately
 arrive at better scheduling decisions.

Recent developments in modern hardware, and in particular modern memory technology, motivate database architects to re-evaluate the entire DBMS systems architecture and in-memory data structures~\cite{10.5555/1325851.1325981, 10.1145/3035918.3054780,DBLP:conf/damon/RenenVL0K19}. Over the years, research in this area has delivered promising propositions, \eg,~\cite{10.1145/2723372.2749441,10.1145/3399666.3399900,DBLP:journals/pvldb/LerschSOL20,10.1145/3183713.3196898}.
 In contrast, we evaluate how far \emph{existing} technology will take us, given careful, cross-cutting engineering.
 %in the system software stack.  

 %In the upcoming section, we discuss steps towards assessing the generalizability of our approach to more complex database engines, in particular, requiring access to the disk. 

%In practice, latencies are far from deterministic, and using cloud services, can be considerable. Performance variance has been recognized to constitute a serious problem in data management community~\cite{DBLP:journals/pvldb/SchadDQ10}

%performance predictability has long since been lamented (over  a decade ago)~\cite{armbrust09abovetheclouds}, with massive use of machine virtualization being one of the root causes~\cite{180314}

%TODO: intrinsische vs.\ systemische Latenz einführen

%Real-time systems community used concept of WCET \dots has also been explored for real-time databases~\cite{DBLP:books/idea/encyclopediaDB2005/Buchmann05}, but is commonly considered to be realistic only for in-memory database enines.

\vspace*{-1em}\section{Experiments}
\label{sec:experiments}

We next describe the setup of our experiments, and then present our results.
 Our Docker image\footnote{Available online from \url{https://github.com/lfd/btw2021}.}, which we describe in Appendix~\ref{sec:appendix-reproducibility}, allows for inspection and reproduction.

\vspace*{-1em}
\paragraph{Database Engine.}
We conduct our experiments with the in-memory database engine DBToaster~\cite{DBLP:journals/vldb/KochAKNNLS14}.
DBToaster can compile SQL queries to C++ code, which we then compile (in a second step) for our target platform. The resulting executable is a single-threaded database engine that incrementally updates a SQL view given a tuple stream.
DBToaster is thus a SQL-to-code compiler, designed to maintain materialised SQL views with low refresh latencies. Typical application scenarios would be in stream processing, such as algorithmic trading, network monitoring, 
or clickstream analysis\footnote{See the project homepage at \url{https://dbtoaster.github.io/home_about.html}, last accessed January 2021.}.
The DBToaster system and its theory have been prominently published (\eg,~\cite{DBLP:journals/vldb/KochAKNNLS14,DBLP:conf/pods/Koch10,DBLP:conf/pods/0001LT16,DBLP:conf/sigmod/NikolicD016}).

We have created our own fork of the DBToaster code base (which is open source), with minor modifications  for our experiments (e.g., buffering measurement data in memory, rather than writing directly to standard output). Our fork  is part of our reproduction package.

\vspace*{-1em}\paragraph{Data and Queries.}
We consider two benchmark scenarios from the DBToaster experiments in~\cite{Koch:183767}.
To be able to discuss the run-time results in greater detail, we focus on only a subset of queries. In particular,
we exclude queries that display a high level of \emph{intrinsic} variability in their latencies,  where the computational effort between tuples can vary greatly, for instance because of nested correlated sub-queries and multi-joins. These queries are \emph{per se} not well-suited for stream processing.
The queries considered by us are listed in Figure~\ref{fig:queries}.

\emph{Finance queries.}
 The queries over financial data process a tuple stream with stock market
activity; we chose three queries which use different relational operators:
 Query \emph{countone}~(C1) is designed by us and serves as a minimal baseline. 
    DBToaster can incrementally evaluate this query with constant-time overhead per tuple. 
 The queries \emph{axfinder}~(AXF) and \emph{pricespread}~(PSP) each compute a join, a selection, aggregation, and in the case of \emph{axfinder} also a group-by on the input stream.
Here, we use the exact same query syntax as in~\cite{Koch:183767}, as DBToaster has certain restrictions (e.g., no LEFT OUTER join). 
To be able to execute these queries on hardware devices with very limited memory, we use a base data set of 100 tuples\footnote{\url{https://github.com/dbtoaster/dbtoaster-experiments-data/blob/master/finance/tiny/finance.csv}}, over which we iterate 5k times, yielding 500k data points. Since the query predicates do not filter on time-stamps, this does not affect query semantics.

\emph{TPC-H queries.}
 We generated TPC-H data with the \emph{dbgen} data generator, set to scale factor~4. We chose the queries~Q6, Q1, and~Q11a (shown in Figure~\ref{fig:queries}) from the DBToaster experiments in~\cite{Koch:183767}.
The queries perform selections, aggregations, and in the case of~Q11a also a join.

\begin{figure}[htb]
\noindent
\begin{minipage}{0.49\linewidth}
\begin{querybox}{C1}
\begin{lstlisting}[style=query]
SELECT count (1) FROM bids;
\end{lstlisting}
\end{querybox}
\begin{querybox}{axfinder}
\begin{lstlisting}[style=query]
SELECT b.broker_id, 
 SUM(a.volume+(-1*b.volume)) AS axfinder
FROM bids b, asks a
WHERE b.broker_id = a.broker_id 
  AND ((a.price+((-1) * b.price)>1000)
  OR (b.price+((-1) * a.price)>1000))
GROUP BY b.broker_id;
\end{lstlisting}
\end{querybox}
\begin{querybox}{TPCH Q11a}
\begin{lstlisting}[style=query]
SELECT ps.partkey,  
  SUM(ps.supplycost * ps.availqty)
  AS query11a
FROM  partsupp ps, supplier s 
WHERE ps.suppkey = s.suppkey 
GROUP BY ps.partkey;
\end{lstlisting}
\end{querybox}
\end{minipage}\hfill
\begin{minipage}{0.49\linewidth}
\begin{querybox}{pricespread}
\begin{lstlisting}[style=query]
SELECT SUM(a.price + (-1*b.price))
       AS psp
FROM bids b, asks a
WHERE (b.volume > 0.0001 *
 (SELECT SUM(b1.volume) FROM bids b1))
  AND (a.volume > 0.0001 *
 (SELECT SUM(a1.volume) FROM asks a1));
\end{lstlisting}
\end{querybox}
\begin{querybox}{TPCH Q6}
\begin{lstlisting}[style=query]
SELECT SUM(l.extendedprice*l.discount)
       AS revenue
FROM lineitem l
WHERE l.shipdate>=DATE('1994-01-01') 
  AND l.shipdate<DATE('1995-01-01')
  AND ( l.discount BETWEEN (0.06-0.01) 
        AND (0.06+0.01) )
  AND l.quantity<24;
\end{lstlisting}
\end{querybox}
\end{minipage}
\begin{querybox}{TPCH Q1}
\begin{lstlisting}[style=query]
SELECT returnflag, linestatus, 
  SUM(quantity) AS sum_qty, SUM(extendedprice) AS sum_base_price,
  SUM(extendedprice*(1-discount)) AS sum_disc_price,
  SUM(extendedprice*(1-discount)*(1+tax)) AS sum_charge,
  AVG(quantity) AS avg_qty, AVG(extendedprice) AS avg_price,
  AVG(discount) AS avg_disc, COUNT(*) AS count_order
FROM lineitem 
WHERE shipdate<=DATE('1997-09-01') 
GROUP BY returnflag, linestatus;
\end{lstlisting}
\end{querybox}
\caption{SQL queries used in the experiments (queries from~\cite{Koch:183767}, with the exception of C1).}\label{fig:queries}
\end{figure}

%\smallskip
%\noindent
\vspace*{-1em}\paragraph{Execution Platforms.}
%\paragraph{Execution platforms.}
For x86 reference measurements, we use a Dell PowerEdge T440.
The T440 is equipped with a single 12 core Intel\textsuperscript{\textregistered\texttrademark} Xeon\textsuperscript{\textregistered\texttrademark} Gold 5118 CPUs and 32~GiB of main memory.
For measurements on Linux, we use kernel version 5.4.38 (vanilla kernel as provided by kernel.org) as baseline, with the Preempt\_RT real-time preemption patch. 
%The kernel is configured as preemptible kernel (CONFIG\_PREEMPT), but does not use hard 
%real-time preemption.

Since delays are caused by parallel access to shared execution units and resources, symmetric multithreading (SMT) is a source of undesired high latencies and noise in real-time systems.
Consequently, we deactivate SMT on our target, in accordance with the original DBToaster experiments. 
Furthermore, we deactivate Intel\textsuperscript{\textregistered\texttrademark} Turbo Boost\textsuperscript{\textregistered\texttrademark}, as sporadic variations of the core frequency result in non-deterministic execution times for identical computational paths.
We configure the CPUs in the highest possible P-State (performance setting) that guarantees a stable core frequency of 2.29~GHz.

For the shielding scenario, we try to remove all operating system noise from the target CPU. The Linux kernel provides multiple mechanisms for this purpose, of which we choose 
CPU namespaces that can be dynamically reconfigured during system operation.\unskip\kern-2pt\footnote{Other mechanisms like CPU isolation at boot-time would provide a slightly higher level of isolation, but must be statically configured at boot-time,
limiting the flexibility of the setup.}

For the partitioned Jailhouse setup, we release one single CPU and 1~GiB of main memory from Linux, and assign them to a new computational domain.
On that domain, we boot the RTEMS + DBToaster binary%
\footnote{Getting DBToaster to run on RTEMS was not straightforward; along our trials, several fixes were proposed to open source systems, such as a decade-old bug revealed in GCC, as well as a bug identified in RTEMS.}
that runs in parallel to Linux.
We use Intel's Cache Allocation Technology (CAT), part of Intel's Resource Director Technology, to partition last-level caches and exclusively assign 5~MiB of Level 3 Cache (L3\$) to the RTEMS + DBToaster domain.
This mitigates noise (cache pollution) of neighboured CPUs, as the L3\$ is shared across all cores~\cite{intel-cat}.

For the ARM reference platform, we use a BeagleBone Black with a single-core Sitara AM3358, a 32 bit ARM Cortex-A8 processor and 512~MiB of main memory.
In contrast to the powerful Intel server CPU, such ARM processors are typically found in embedded or industrial applications.
We boot the RTEMS + DBToaster application directly on bare-metal.

\paragraph{Methodology.}
DBToaster logs a time-stamp for every \(N\) input tuples processed.
This allows us to compute the \emph{latency per \(N\) input tuples processed}, averaged over \(N\) tuples. While averaging is a sensible
and established choice for throughput measurements to minimise overhead of the measurement intervention, we are interested in a precise characterisation of system noise
vs.\ intrinsic variation of the core processing code, and therefore resort
to measuring processing times on a \emph{per-tuple} basis (\(N=1\)). 

We distinguish between two units of measurements:
(1)~time stamps obtained by the standard POSIX API (\texttt{clock\_gettime} with \texttt{CLOCK\_MONOTONIC}). This allows for nanosecond resolution, but also inflicts considerable overheads in the microsecond range, and introduces a noise level that is on par with the processing time proper for some of the simpler queries. Therefore, we extend DBToaster with the optional capability of (2) using x86 time stamp counter (TSC) ticks. While there are several problems and pitfalls associated with using the TSC on SMP configurations, and while the
obtained measurement values cannot be converted to walltime without further ado~\cite{mauerer2010professional}, TSCs are one of the highest-precision clock sources available on x86 hardware,
and can be read from userspace 
%applications 
without %necessitating
transition to kernelmode.\unskip\kern-2pt\footnote{Using a high-resolution, low overhead time source is not 
necessary on our ARM reference platform because the time required to obtain
a time stamp is negligible in comparison to the average processor performance, and our operating system has a flat memory and privilege model -- that is, there is no distinction between kernel- and usermode on our near-bare-metal measurements on this platform.}

%\begin{figure}
%    \centering\TODO{Edson to provide methodology image}
%    \includegraphics{}
%    \caption{fig:repro}\label{fig:methodology}
%\end{figure}

%We forked the DBToaster GitHub repository\footnote{\url{https://github.com/dbtoaster/dbtoaster-backend}, commit {\tt 3b17644}}. 
As is a standard approach in settings like ours~\cite{DBLP:journals/arc/BuchmannL01},
we start measuring time once the input is in memory. In particular, we pre-load all tuples prior to stream processing, to exclude noise caused by I/O.
Of course, in any real-world setting, the tuples would be read over peripheral communication channels, such as ethernet.
To further avoid noise in our measurements, we have modified the code generated by DBToaster such that these time-stamps are cached in memory during query evaluation, in a pre-allocated array, rather than being  continuously written to the standard output console.
%This is only a minor change at that, to the DBToaster engine.
%The measurement process is illustrated in Figure~\ref{fig:methodology}.

%In compiling SQL queries with DBToaster, we did not use the compilation flags recommended in~\cite{Koch:183767}, as we found that they are no longer supported.

\vspace*{-1em}\subparagraph{Simulating tenant load.}
We simulate further tenants executing on the same system using the standard utility \href{https://wiki.ubuntu.com/Kernel/Reference/stress-ng}{\texttt{stress-ng}}, running 6 synthetic workloads.\unskip\kern-2pt\footnote{(1)~Binary search on a sorted array (exercises random memory access and processor caches), 
(2)~matrix multiplication (to stress memory cache and floating point units), (3)~compressing/decompressing random data (exercising CPU, cache, and memory), 
(4)~randomly spread memory read and writes (to thrash the CPU cache), (5)~sequential, random and memory mapped read/write operations (to exercise the I/O subsystem), and (6)~timer interrupts at the frequency of 1~MHz (to induce continuous kernel/userspace transition due to interrupt handling).}
In Figure~\ref{fig:software-stack}, we depict \cmd{stress-ng} running as additional load on the CPUs that are annotated with ``L''.

\begin{figure}[htb]
    \input{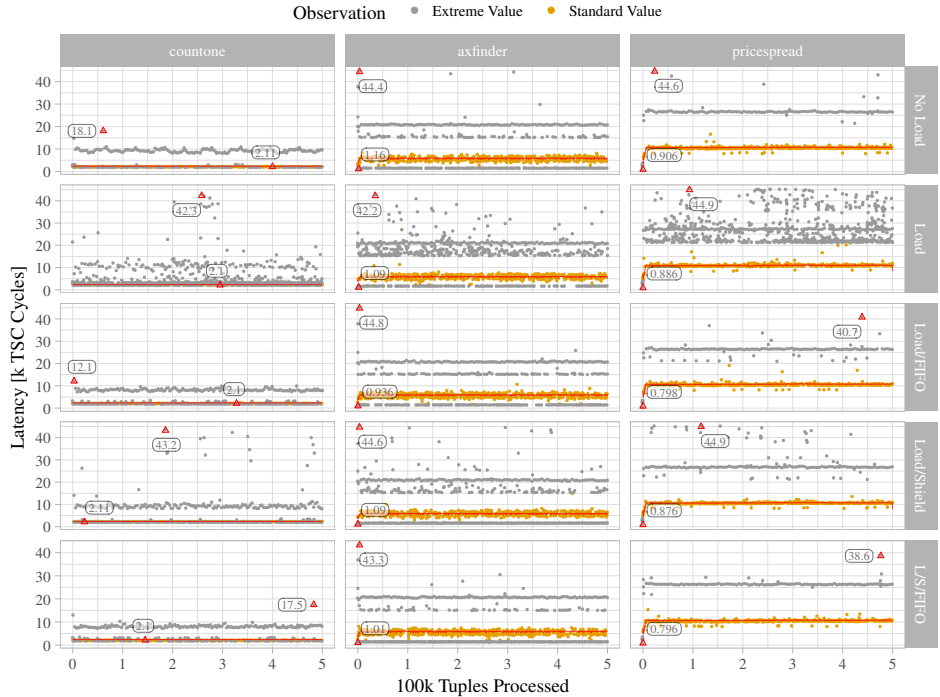}\vspace*{-2.25em}
    \caption{Latency time series for finance queries on x86, using
    the high-speed time stamp counter (TSC).  }\label{fig:latencies_finance_x86_tsc}
\end{figure}

\subsection{Results}
\subsubsection{Noise and Determinism: Finance Queries}
We begin our discussion of results for the
finance queries.
The time series in Figure~\ref{fig:latencies_finance_x86_tsc} show observed latencies for processing each out of 500k input tuples.
Red, labelled triangles mark the minima and maxima.
Since almost all measured
values fall into a comparatively narrow standard range, which would lead to massive
over-plotting and loss of information, we colour all ``extreme'' measurement points that fall in the bottom 0.05\% percentile, or that exceed the 99.95\% percentile, in grey. We consider all other data points (marked ochre) as the normal observations. Note that such outliers have no noticeable influence when it comes to performance measurements, which usually concern query throughput, based on temporal averages, but are of paramount importance for real-time,
bounded latency scenarios. For instance, the experiments in~\cite{DBLP:journals/vldb/KochAKNNLS14} consider the query refresh times for processing batches of 1,000 tuples, and we compute a sliding mean window
over 1,000 tuples as a consistency check; the resulting red line nicely reproduces the original DBToaster experiments~\cite{DBLP:journals/vldb/KochAKNNLS14}. 

Each subplot of a given column corresponds to one system software stack scenario from Section~\ref{sec:prelims}.
Almost all latencies are centred around the sliding mean value. However, a few outliers exceed the mean by a factor of about four.\unskip\kern-2pt\footnote{In the scenario discussed in this paper, %the essential criterion 
%for real-time performance is 
the \emph{maximum observed latency} is essential. Exceeding a threshold in industrial control scenarios might have severe consequences, from lost capital over destroyed machinery to bodily harm or loss of life, which can
never be compensated by the fact the this does not happen \emph{on average}.}

We have also tracked the average performance 
of the simulated other tenants, and found that it was essentially identical regardless of the measurement
setup, which shows that improved determinism for a given workload does not necessarily
decrease average throughput for non-real-time loads. Detailed data are available in the reproduction package.

While we consider queries of different intrinsic complexity, there is no
direct relation between query complexity and noise -- however, there \emph{is} a
relation between query complexity and \emph{average} performance, as visible in the 
increasing latencies of the red line from left (simpler query) to right (more complicated query). 

Query \emph{countone} merely counts the number of input tuples processed so far (and is refreshed for each input tuple), whereas the other finance queries compute joins. As can be expected, the average latency for \emph{countone} is distinctly lower. For the other queries,
we can observe densely populated discrete "horizontal bands" that group the majority of all observed values. They correspond, based on an analysis of profiling data, to the main execution paths
taken by the DBToaster-generated code (two main execution paths are a consequence of the
``orderbook adapter'' that distinguishes between the two types of input data, bids and asks). Also, when DBToaster-internal dynamic data structures grow in size (such as when buffering tuples for computing hash joins), additional DBToaster-intrinsic latencies incur.

The vertical spread of observations around these bands is an obvious visual noise measure. By comparing against the ``Load'' scenario, it is visually apparent that the different isolation mechanisms %considered in this paper
substantially reduce the observed jitter, typically to the
level of an otherwise unloaded system. The strongest form of isolation, CPU shielding plus realtime scheduling (L/S/FIFO), produces latency distributions that are not only comparable, but even better in terms of maximum
values than in the ``No Load'' scenario.
%, which is undemanding from a systems point of view, but the usual scenario considered in throughput measurements. 
The amount of noise decreases
in order Load/Shield, Load/FIFO, and Load/Shield/FIFO. It might surprise
%at a first glance 
that a shielded CPU performs worse than a CPU with additional load,
but with a real-time prioritised task of interest. Recall that there is a complex
interaction of kernel features as outlined in Sec.~\ref{sec:exp_conditions}, and that, for
instance, a larger set or possible preemption points, together with delayed kernel
administrative work in a shielded scenario, may well compensate the advantages gained by
exclusive CPU access.

%The financial queries operate over  homogeneous data, and even though two of the queries compute a two-way join, they are based on simple constructions. 
While the
measurements show a noticeable reduction of noise when using more advanced isolation techniques, the reduction of maximal latencies comprises only a factor of two.
%We next contrast this with our discussion of the TPC-H queries.
%, which are more sensitive towards noise.
%, which are based on a more complex database schema.

%\begin{table}[ht]
%\centering%\label{tab:tpch_tsc}
%\caption{Latency variance for TPC-H queries. \TODO{Regenerate}}
%\begin{tabular}{rllrrrrrr}
%  \hline
% & compare & query & min & q1 & mean & median & q3 & max \\ 
%  \hline
%1 & No Load & 1 &   0 & 42.00 & 247.01 & 92.00 & 168.00 & 147122 \\ 
%  2 & Load & 1 &   0 & 38.00 & 9123.90 & 82.00 & 142.00 & 117059556 \\ 
%  3 & Load/FIFO & 1 &   0 & 40.00 & 171.66 & 84.00 & 146.00 & 213294 \\ 
%  4 & Load/Shield & 1 &   0 & 42.00 & 176.11 & 88.00 & 154.00 & 308906 \\ 
%  5 & Load/Shield+FIFO & 1 &   0 & 0.00 & 0.00 & 0.00 & 0.00 &   0 \\
% % 6 & No Load & 6 &   0 & 30.00 & 122.29 & 66.00 & 116.00 & 181780 \\
% %5 7 & Load & 6 &   0 & 30.00 & 8382.46 & 66.00 & 116.00 & 149178302 \\ 
% % 8 & Load/FIFO & 6 &   0 & 28.00 & 118.40 & 62.00 & 108.00 & 181898 \\ 
%%  9 & Load/Shield & 6 &   0 & 30.00 & 117.31 & 62.00 & 110.00 & 303806 \\ 
%%  10 & Load/Shield+FIFO & 6 &   0 & 0.00 & 0.00 & 0.00 & 0.00 &   0 \\ 
%%  11 & No Load & 11a &   0 & 32.00 & 198.25 & 86.00 & 206.00 & 21209986 \\ 
%%  12 & Load & 11a &   0 & 28.00 & 13510.60 & 74.00 & 194.00 & 1034330146 \\ 
%%  13 & Load/FIFO & 11a &   0 & 28.00 & 176.77 & 70.00 & 188.00 & 5384582 \\ 
%%  14 & Load/Shield & 11a &   0 & 28.00 & 174.05 & 70.00 & 186.00 & 1852986 \\ 
%%  15 & Load/Shield+FIFO & 11a &   0 & 0.00 & 0.00 & 0.00 & 0.00 &   0 \\ 
%   \hline
%\end{tabular}
%\label{tab:tpch_tsc}
%\end{table}

\subsubsection{Noise and Determinism: TPC-H Queries}
\begin{figure}[tb]
    \input{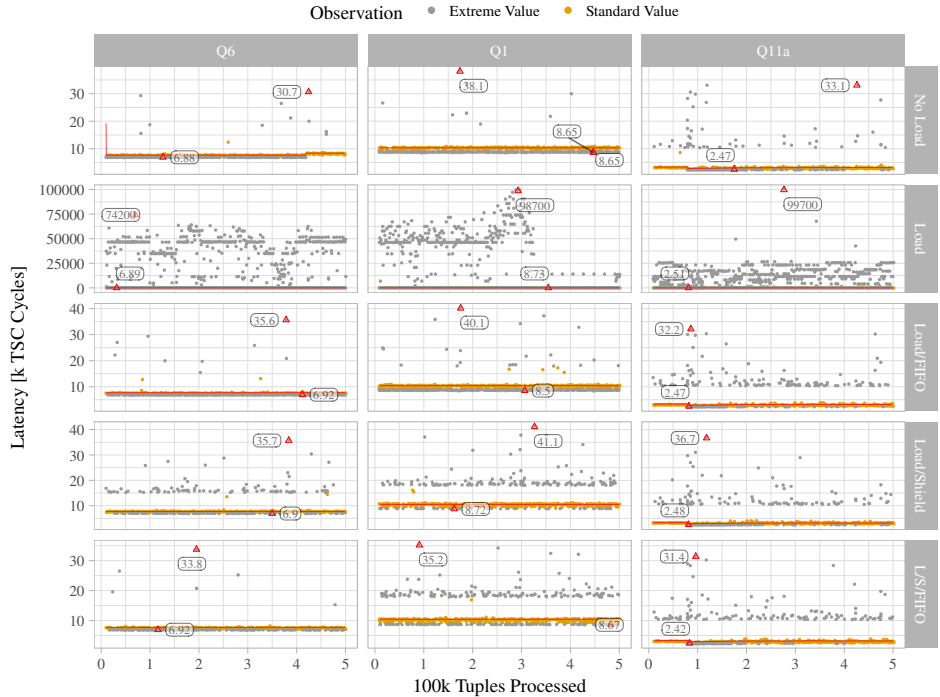}\vspace*{-2.25em}
    \caption{Latency time series for TPC-H queries on x86, using the high-speed time stamp counter (TSC).}\label{fig:latencies_tpch_x86_tsc}
\end{figure}

The latency measurement results for the TPC-H queries 
are shown in Figure~\ref{fig:latencies_tpch_x86_tsc}.
%, using the same conventions as before. 
While the general observations are identical -- measured
values concentrate around a few horizontal ``bands'',\footnote{Notice that such bands may
also be caused by  system effects, and are then not necessarily present in all measurement
combinations: For instance, the Load/Shield scenario in Figure~\ref{fig:latencies_tpch_x86_tsc}
contains a band that disappears when FIFO scheduling is  activated. Bands
present in all scenarios are typically, but not necessarily, caused by the payload
software. Such detail observations are not possible
in %typical standard 
measurements that average over %multiple 
observations.} and noise decreases
with the various forms of systemic isolation -- the behaviour of the queries under
high load differs considerably from the ``No load'' and isolated case: The difference
in maximal latencies comprises more than three decimal orders of magnitude (observe the different scales of the vertical axes in the plots), and a similar 
statement can be made for the width of the spread around the running mean value,
the latter again plotted with a red line. While such high variance has grave consequences
for real-time systems, it is not even observable when throughput
measurements are averaged. 

\begin{figure}[htb]
    \input{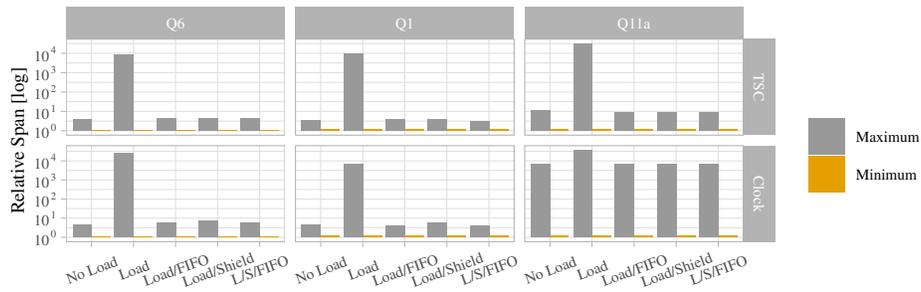}\vspace*{-3.0em}
    \caption{Span of observations relative to median query latency of the TPC-H queries in Figure~\ref{fig:latencies_tpch_x86_tsc}. Clock: Measured using \texttt{clock\_gettime}. TSC: Measured using the high-resolution time stamp counter.}
    \label{fig:span_tpch}
\end{figure}

\newcommand{\med}{\ensuremath{\operatorname{med}}}
So far, we have relied on visual means for characterising noise. For a
quantitative measure, consider the set of observed latencies \(\{\Delta t_{i}\}\)
(each data point in Fig.~\ref{fig:latencies_tpch_x86_tsc} corresponds to one
value of \(\Delta t_{i}\)). 
While we have focused on x86 so far, we
will also consider ARM-based systems below. These platforms vary widely in their
performance,
and absolute values consequently require interpretation. It is therefore
pertinent to consider \emph{relative} deviations from the expected response time,
which allows us to compare %the level of determinism. %present 
across platforms. 

To this end, we define \emph{spreads}, which are \emph{not} influenced by the absolute
processing speed. The \emph{maximum spread} is given by
\(\max(\{\Delta t_{i}\})/\med(\{\Delta t_{i}\})\), and \emph{minimum spread}
by \(\med(\{\Delta t_{i}\})/\min(\{\Delta t_{i}\})\), where \(\med(\cdot)\) denotes
the median of the argument set. The quantities characterise the system-global relative
span between a ``typical'' observed value, and the most extreme outliers
in both directions. The 
results shown in the row labelled ``TSC'' of Fig.~\ref{fig:span_tpch} %(the other row is covered further below)
quantitatively underlines this: Spread in the
``Load'' scenario typically exceeds the spread in the isolation scenarios by 
orders of magnitude. 
The differences between the various isolation scenarios
(and the ``No Load'' case) are much less pronounced, but can still encompass
a factor of two or three (note the log-transformation of the  \(y\) axis).

Recall that we distinguish between two units of measurements for latencies,
(1)~wallclock time in nanoseconds, and (2)~x86 time-stamp counter ticks. 
The row labeled ``Clock'' of Figure~\ref{fig:span_tpch} highlights another issue 
related to this fact that
is mostly technical, but nonetheless requires careful consideration: how to
perform the measurement itself. It shows the relative span for the
\emph{identical} measurement as considered in the other row, but
this time using per-tuple latency measurements based on the POSIX API call
\texttt{clock\_gettime} offered as service by the Linux kernel (and often
replaced by the lower-precision variant \texttt{gettimeofday}, 
in a good fraction of published performance measurements). Especially
the maximum span can differ considerably among measurement variants.
For TPC-H query~11a, it is even the major source of noise, as the right part of
Fig.~\ref{fig:span_tpch} shows. 

\begin{figure}[htb]
    \input{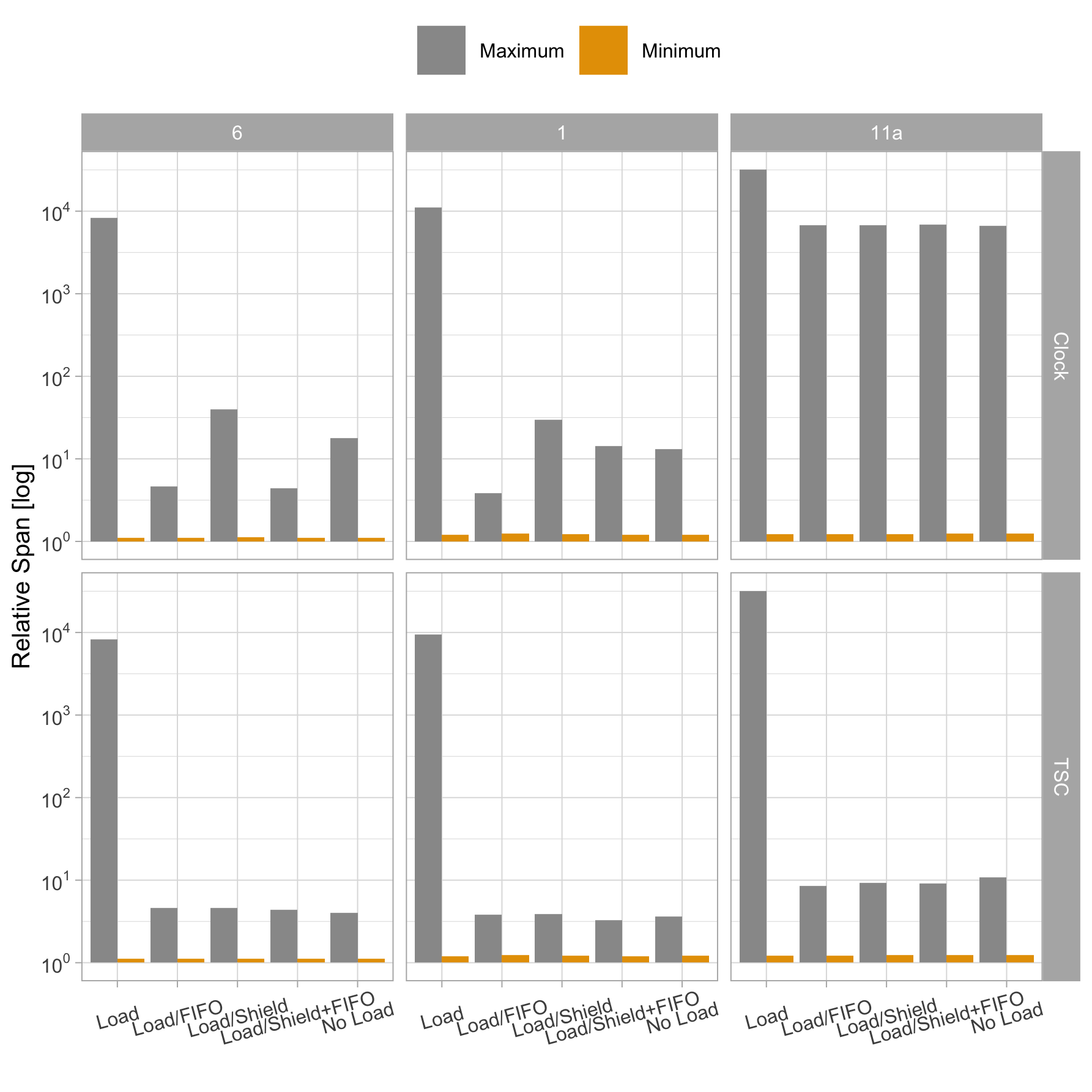}\vspace*{-2.25em}
    \caption{Latency time series for TPC-H queries on x86, 
    using a kernel-provided clock.}\label{fig:latencies_tpch_x86_clock}
\end{figure}

Fig.~\ref{fig:latencies_tpch_x86_clock}
illustrates, for a subset of the isolation mechanisms, the increase in 
spread and noise distribution for clock-based measurements. It particularly
highlights that even the mean throughput value (red line) is substantially influenced by the increased overhead.

\subsection{The Role of CPU Noise}
To a major extent, the previous experiments concern the control
of noise introduced by the operating system and the presence of
other tasks that compete for CPU time and other shared 
resources. Especially the scenario using CPU isolation, combined
with a real-time scheduling policy, eliminates a substantial
fraction of this noise. We now question how much of the remaining noise is caused by the executing
CPU itself, and can thus be seen as an effective lower bound
on any systemic noise.
%, respectively serve as a characterisation of the intrinsic query noise.

We thus run our binaries as close to the
bare-metal as possible, which %by design 
reduces OS overhead.
We perform these measurements on an ARM processor
that we deem powerful enough to execute reasonable database operations,
but that uses substantially fewer performance optimisations
than x86 server-class CPUs (and, thus, suffers from less intrinsic noise).
Our choice for an ARM CPU is not just driven by simplicity, though: Processors
of this type are the most frequent choice in embedded systems and IoT
devices, where low latency data processing is a common requirement (for instance,
think of sensor-based systems that derive action decisions by combining
previously measured values stored in a database with current data points).
Our measurements are therefore representative for this large class of
systems that we expect will gain even more importance in future applications.
Of course, measurements on CPUs with drastically different capabilities cannot be 
directly compared, and this is not our desire: Instead, it is important to consider the \emph{relative} difference between average and maximal latencies, and the span within measurements, as discussed below.

\begin{figure}[htb]
    \input{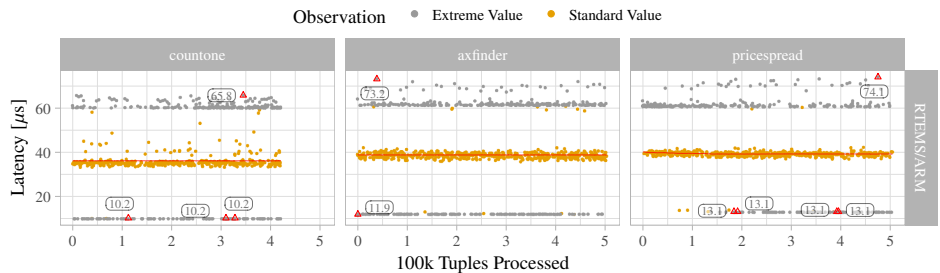}\vspace*{-2.25em}
    \caption{Latency time series for finance queries on an ARM system
    (BeagleBone Black) using RTEMS. Red, labelled triangles represent minima and maxima (not necessarily unique).
    }\label{fig:latencies_bbb}
\end{figure}

\begin{figure}[htb]
    \input{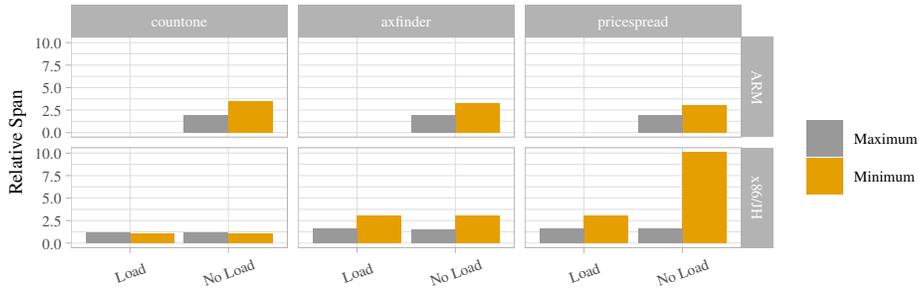}\vspace*{-3.5em}
    \caption{Span of observations relative to median query latency of finance queries on bare-metal, on a high-end (x86) and low-end (ARM) CPU.}\label{fig:span_bare}
\end{figure}

 Fig.~\ref{fig:latencies_bbb} shows latency time series for three finance queries. Again, observations centre
around horizontal bands induced by the main execution paths, but the overall
jitter is limited. The reduction compared to Jailhouse on x86 is quantified in
Fig.~\ref{fig:span_bare}.

\begin{figure}[htbp]
    \input{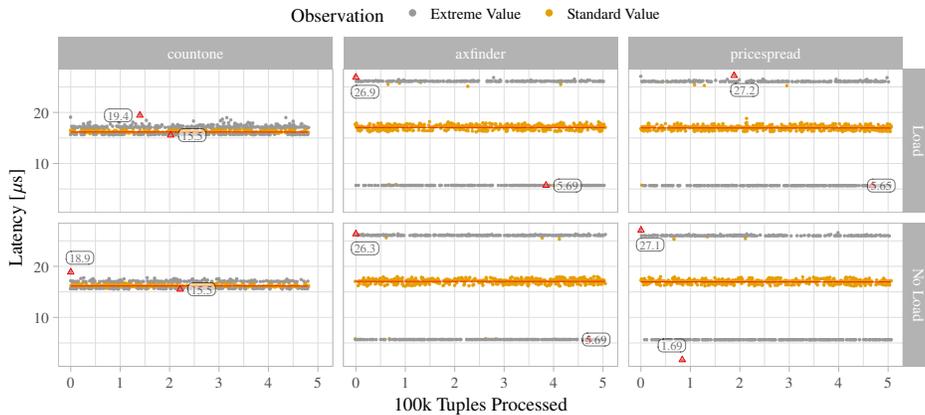}\vspace*{-2.5em}
    \caption{Latency time series for finance queries on an RTEMS-based near bare-metal
    CPU provided by the Jailhouse hypervisor.% Red, labelled triangles represent minima and maxima.
    }\label{fig:latencies_jh_x86}
\end{figure}

The summary for a second set of measurements shown in the bottom part of
Fig.~\ref{fig:span_bare} represents bare-metal results obtained on the x86 CPU,
but this time driven by an RTEMS kernel running inside a Jailhouse cell.
Since the system is equipped with a total of 12 cores (compared to the
single-core ARM), and only one of the cores is needed to run the
database workload, we extend the measurement with an additional aspect
that quantifies the aptitude of the setup to decouple latency-critical
database operations performed by one tenant from other, perhaps
throughput-oriented operations performed by other tenants. The spread
is, as Fig.~\ref{fig:span_bare} shows, almost identical between the
scenarios. This is also reflected in the time series shown in
Fig.~\ref{fig:latencies_jh_x86}, which demonstrates that the results of the
two configurations do not deviate in any substantial way. Since the isolation
provided by Jailhouse does not only address execution timing, but also extends
to other security and privacy related aspects of database workload
processing, 
we deem this configuration a suitable basis for 
multi-tenant systems with strong separation guarantees.

%\TODO{(1) High-prio: Show that the workload from other tenants is not suffering.
%(2) Lower-prio: State the sizes of the binaries for the software systems stack, and how %small we can go. This can also be inlined in the text.
%(3)  Refer to an online appendix that contains charts for further queries.}

\section{Consequences}
\label{sec:discussion}

Our experimental results show that -- at least for our use-case of an in-memory database engine --  building a database stack on a plain-vanilla Linux with custom settings can already compete with running close to bare-metal on the hardware.
In our experiments we reached a state where the major source of noise turned out to be the interruptions to measure time (and noise) itself --  any other sources for system-software induced jitter had been eradicated.
In the following, we discuss some of these results in a broader context.

\vspace*{-1em}\paragraph{There's life in the old dogs, yet.}
Our results suggest that it may not be necessary to design dedicated DB-aware operating systems \emph{from scratch}. Rather, a prudent strategy to extend and enhance existing systems selectively may pay off equally well, and provide faster results.
This is a recurring experience in the systems community: About a decade ago, the upcoming "multicore challenge" was supposed to render existing system-software designs unviable, and new kernel designs were deemed necessary~\cite{boyd-wickizer:08:osdi,wentzlaff:09:acmosr,baumann:09:sosp}.
It later turned out that existing system software could be scaled-up almost equally well by a systematic examination of their bottlenecks, which then could be fixed by employing standard techniques of parallel programming combined with \emph{a few} novel abstractions~\cite{boyd-wickizer:10:osdi}.
In a similar run--analyse--fix-approach, we have shown that existing system software, such as Linux, might just be \emph{good enough} for many more database use cases, given proper configuration and adaptation.
Of course, this does not invalidate the ongoing research on novel operating systems customised for database engines.
Instead, the lesson to be learned here is that studying the \emph{actual} reasons behind noise observed with existing operating systems is important.
Only if we can pinpoint and understand the root causes, can we think of innovative solutions to these problems.

\vspace*{-1em}\paragraph{The cure might come by foreigners.}
Basically all of the measures we applied have originally been introduced into Linux to improve determinism and worst-case latencies not in database engines, but for the domain of embedded real-time systems: SCHED\_FIFO, CPU shields, interrupt redirection, and PREEMPT\_RT were developed and introduced to make Linux a suitable platform for mixed-criticality~\cite{vestal:07:rtss,burns:18:acmcs} workloads in the real-time domain.
This is also the main motivation behind partitioning hypervisors, such as Jailhouse.  
In our understanding, multi-tenant database engines that need to provide isolation and a guaranteed quality of service could (and should) be considered as (soft) real-time systems. 
Hence, operating-system solutions originally developed for the real-time domain might be a promising solution vector for the development of time-critical database systems.
This underlines the necessity to rejoin the database and system communities.

\vspace*{-1em}\paragraph{Many challenges remain.}
Even though our results are promising, it should be pointed out that we eradicated only the CPU noise, and deliberately ignored I/O noise.
Compared to the (narrow) case of a pure in-memory database, disk-based or otherwise
I/O-intensive database stacks can be treated using the same measures as
employed in this paper, but would face different challenges.
In fact, most of Linux's built-in real-time measures do not interact well with the I/O subsystem without further ado; it is often assumed that the time-critical part can be
decoupled from all I/O activities.
While we have not yet examined disk-based database engines in this respect, we
expect this to become a larger challenge that probably requires more invasive changes
to the existing software stack. Direct I/O~\cite{peter:14:osdi} might be a promising
way to approach this. Likewise, low-latency~\cite{Lee:2019} or deterministic~\cite{10.1145/2815400.2815421,Swaminathan:2003} I/O scheduling have received
a fair amount of attention outside the database
community. External input via networks must consider additional stochastic parameters
(\eg, unpredictable arrival times of data packets) that add to the complexity of the
investigation. Real-time~\cite{Kiszka:2005}, or time-sensitive networking,
and userland-based low-latency interaction with networking hardware can also be applied in database use-cases, albeit details must be left to future work.

\vspace*{-1em}\paragraph{What's next.}
Consider, as a specific and current example, how in-memory
database engines can be extended with disk support -- as, for
instance, happens in the extension of Hyper to Umbra, where
the authors propose to use SSDs for storage~\cite{DBLP:conf/cidr/NeumannF20}. Especially
parallel combinations of multiple SSDs promise RAM-like access performance.

However, parallel SSDs must be managed and driven. Database engines frequently aim at
controlling block devices (at least to schedule access) from
userland, since they have more complete usage pattern information than the OS. Yet this
approach inevitably suffers from (at least) the amount of noise we have observed
in our measurements, and advanced functionalities like RAID require substantial
engineering effort. Operating systems provide such services as a commodity, but 
 lack integration with the database query optimiser and its cost model.
Additionally,
operating systems are commonly optimised for throughout, so considerable tail latencies can
be expected without adaptations. However, we are optimistic that
moderate extensions of existing kernel mechanisms will combine the benefits
of already existing infrastructure with little noise.
This is 
important since increased determinism is beneficial to finding 
optimal query plans. 

For all of the challenges listed above, we are optimistic that the required changes will be comparatively small compared to
developing a new operating system from scratch.

%To avoid pronounced tail latencies as for pure memory access, because locality increasing %mechanisms like system partitioning into NUMA nodes do not apply likewise
%to I/O with block devices, and because the operating systems used as
%basis for most databases are optimised for throughput, not determinism.

%Traditionell: DB verwalten Blockgeraete selbst, weil Scheduling + Queryplan bekannt => Optimioerung Kopfbewegung rotierender Rost aus DB effizienter als aus OS

%erster Schritt von mm-based auf disk-based mit Einsatz von SSDs vorgeschlagen

%SSD => Hoeherer Bandbreite => (total vielversprechend:) Umbra mit near-memory-performance => 
%Mehrere SSD wollen getrieben werden => Wenn aus userland, dann 
%Einfluss von systemischem Rauschen mindestens genauso hoch wie
%bei unseren Load-Experimenten 
%=> Wenn Determinismus (neben Durchsatz
%erweunscht), Kooperation mit OS-Kern, statt gegeneinander zu arbeiten
%=> Identify, analyse, enhance-Ansatz auf dieses Gebiet anwenden

\section{Conclusion}
\label{sec:conclusion}

We have shown that proper use of standard mechanisms of full-featured OSes
can achieve database query latencies comparable to running an in-memory database engine directly on raw hardware. 
We reach a point where measuring time becomes the largest source of noise.
%
%Ultimately, we reduce noise rooted in CPU sharing. 
%One next logical move is from single-threaded to multi-threaded database engines, and to consider transactional workloads. A particularly challenging 
%task is to address I/O, and thus, step from in-memory to disk-based database engines. While we expect a considerable increase in problem complexity (disk access is a known game changer in controlling latencies for real-time databases~\cite{DBLP:books/idea/encyclopediaDB2005/Buchmann05}), we are confident that careful cross-cutting engineering is again key to success.
%
%\begin{samepage}
By addressing challenges beyond CPU noise, we plan to bridge to the domain of real-time systems, and leverage techniques established for mixed-criticality systems which we apply to the database domain.
After all, the underlying ideas match our scenario: One workload (the database  engine) is to be shielded, without impairing the %execution of the
other workloads.
We are confident that the respective research communities will enjoy many mutual
benefits.
%\end{samepage}

\section{Appendix: Reproduction Package}
\label{sec:appendix}

\begin{figure}[tbp]
\centering
\resizebox{0.7\textwidth}{!}{
% https://latexcolor.com/
\definecolor{magnolia}{rgb}{0.97, 0.96, 1.0}
\definecolor{mediumchampagne}{rgb}{0.95, 0.9, 0.67}
\definecolor{linen}{rgb}{0.98, 0.94, 0.9}
\definecolor{amber}{rgb}{1.0, 0.75, 0.0}

% \tikzsetnextfilename{software-stack-noload}
\begin{tikzpicture}[remember picture,
    % framed,
    % background rectangle/.style={thin,draw=black,fill=white},
    system/.style={draw,thin,inner sep=0pt,rectangle,rounded corners=1pt,color=black,fill=lightgray!20,font=\ttfamily},
    group/.style={draw,thin,inner sep=0pt,rectangle,rounded corners=1pt,color=black,fill=white!20,font=\ttfamily},
    numbers/.style={draw,thin,circle,color=black,fill=yellow!20,font=\ttfamily},
    empty/.style={draw,thin,circle,color=white,fill=white!20,font=\ttfamily},
    label/.style={font=\ttfamily},
]

\draw node[system,minimum width=4cm, minimum height=2cm, align=left] (docker) {Docker Container};
%\node[label,anchor=north west,align=left, inner sep=5pt] at (docker.north west) {Docker\\Container};

\node[system,above= of docker,minimum width=2.7cm, minimum height=1cm,align=left] (git) {Public Git\\Repository};
\node[system,right= 0.2cm of git,minimum width=2cm, minimum height=1cm] (patches) {Patches};
\node[system,left= 0.2cm of git,minimum width=2cm, minimum height=1cm] (binaries) {Binaries};

\node[system,left = of docker,minimum width=1.5cm, minimum height=1cm, align=left] (recipe) {Build \\ Recipe};

\node[group, below = of docker,minimum width=6cm, minimum height=1.8cm] (tarball) {
    \begin{tikzpicture}[remember picture]
        \node[system, minimum width=1.5cm, minimum height=0.8cm,align=left] (queries) {Compiled\\Queries};
        \node[system, left = 0.2cm of queries,minimum width=1.8cm, minimum height=0.8cm] (dispatcher) {Dispatcher};
        \node[system, left = 0.2cm of dispatcher,minimum width=1.8cm, minimum height=0.8cm] (evaluation) {Evaluation};
    \end{tikzpicture}
};
\node[label,anchor=north west,align=left, inner sep=5pt] at (tarball.north west) {Tarball};

\node[system,right = 1.3cm of docker,yshift=15pt,minimum width=2.2cm, minimum height=1cm] (measureA) {};
\node[label,anchor=south,align=center, inner sep=5pt] at (measureA.south) {Measurement A \\ (x86)};

\node[system,below = of measureA,minimum width=2.2cm, minimum height=1cm] (resultsA) {Results A};
\node[system,below = of resultsA,minimum width=2.2cm, minimum height=1cm] (chartsA) {Charts A};

\node[system,right = 10pt of measureA,minimum width=2.2cm, minimum height=1cm] (measureB) {};
\node[label,anchor=south,align=center, inner sep=5pt] at (measureB.south) {Measurement B \\ (ARM)};

\node[system,below = of measureB,minimum width=2.2cm, minimum height=1cm] (resultsB) {Results B};
\node[system,below = of resultsB,minimum width=2.2cm, minimum height=1cm] (chartsB) {Charts B};

% edges w/ one line
\draw[black,-{Latex[length=2mm, width=5mm]},line width=1pt] (binaries.south) to[out=270, in=90] (docker.north);
\draw[black,-{Latex[length=2mm, width=5mm]},line width=1pt] (git.south) to[out=270, in=90] (docker.north);
\draw[black,-{Latex[length=2mm, width=5mm]},line width=1pt] (patches.south) to[out=270, in=90] (docker.north);
\draw[black,dashed,-{Latex[length=2mm, width=5mm]},line width=1pt] (tarball.south) -| (0,-4) -| (3.15,-4) -| (3.15,1.5) -| (measureA.north);
\draw[black,dashed,-{Latex[length=2mm, width=5mm]},line width=1pt] (tarball.south) -| (0,-4) -| (3.15,-4) -| (3.15,1.5) -| (measureB.north);

% edges w/ double lines
\draw[black,-{Latex[length=2mm, width=5mm]},double,double distance=2pt,line width=1pt] (recipe.east) to[out=0, in=180] (docker.west);
\draw[black,-{Latex[length=2mm, width=5mm]},double,double distance=2pt,line width=1pt] (docker.south) to[out=270, in=90] (tarball.north);
\draw[black,-{Latex[length=2mm, width=5mm]},double,double distance=2pt,line width=1pt] (measureA.south) to[out=270, in=90] (resultsA.north);
\draw[black,-{Latex[length=2mm, width=5mm]},double,double distance=2pt,line width=1pt] (resultsA.south) to[out=270, in=90] (chartsA.north);
\draw[black,-{Latex[length=2mm, width=5mm]},double,double distance=2pt,line width=1pt] (measureB.south) to[out=270, in=90] (resultsB.north);
\draw[black,-{Latex[length=2mm, width=5mm]},double,double distance=2pt,line width=1pt] (resultsB.south) to[out=270, in=90] (chartsB.north);

% Numbers
% \node[numbers,right = 3pt of recipe,yshift=13pt] (recipeLabel) {\Large 1};
% \node[numbers,above = 6pt of tarball,xshift=15pt] (tarballLabel) {\Large 2};
% \node[numbers,above = 0.55cm of measureA,xshift=15pt] (tarballLabel) {\Large 3};
% \node[numbers,below = 6pt of measureA,xshift=15pt] (tarballLabel) {\Large 4};
% \node[numbers,below = 6pt of resultsA,xshift=15pt] (tarballLabel) {\Large 5};

% Legend
\node[label,right = 0.2cm of patches, yshift=10pt] (l1) {A $\longrightarrow$ B, B integrates A};
\node[label,right = 0.2cm of patches, yshift=2pt] (l2) {A $\Longrightarrow$ B, B is produced by A};

% Temporal Flow legend
\node[empty, below = -0.7pt of l2, xshift=-60pt, inner sep=2pt] (al3) {};
\node[empty, right = 1.0cm of al3, inner sep=2pt] (bl3) {};
%\draw[black,dashed,-{Latex[length=0.0001mm, width=0.0001mm]},line width=1pt] (al3.east) to (bl3.west);
\draw[black,dashed,-{Latex[length=2mm, width=5mm]},line width=1pt] (al3.east) to (bl3.west);
\node[label,right = 0pt of bl3, yshift=0pt] (cl3) {Temporal flow};
% \node[label,right = -8pt of bl3, yshift=-2pt] (cl3) {,};

\end{tikzpicture}
}
\caption{Components and workflow of the reproduction package.}
\label{fig:repro}
\end{figure}
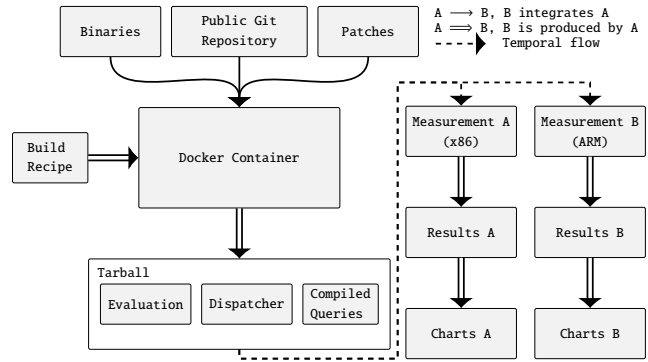

%\subsection{Reproduction Package}
\label{sec:appendix-reproducibility}\looseness=-1
Our publicly available reproduction package is based on Docker.\unskip\kern-2pt\footnote{Available online from \url{https://github.com/lfd/btw2021}.}
The process is illustrated in Figure~\ref{fig:repro}: A Docker build recipe produces the docker container, and
scripts that run therein produce a tarball with executables for all measurements
in this paper. By transferring this tarball to a target, the experiments
can be automatically executed, and charts generated.
 
\looseness=-1
We use binary sources for distribution-level software, 
and build other components (DBToaster, embedded compilers, RTEMS board support packages, \dots) from source using the latest released state, augmented with
local patches, to address issues found during this work that relate to the RTEMS kernel, the
Jailhouse hypervisor, the GNU C compiler, and DBToaster itself (we include patches 
as explicit diff files to make any deviations from  upstream sources explicit without relying on git history inspection). 
Additionally, we do not rely on the long-term
availability of external sources by providing a pre-built Docker image. It contains
all sources and dependencies, and enables re-building the exact same binaries from source that
we use for the measurements (of course, our peers may choose to build the Docker
image from scratch, depending on the latest binaries).
%Document changes to software consistently (following Linux
%kernel development standard), regardless of upstream practice

Finally, we provide all raw measurement results for all system combinations
considered in the paper, and all post-processing scripts to evaluate
and visualise the data.

%\subsection{Benchmark Queries}
%\label{sec:appendix-queries}
%The SQL queries used in the benchmarks are shown in Fig.~\ref{fig:queries}. 
%The financial data scenario is borrowed from the DBToaster
%experiments~\cite{DBLP:journals/vldb/KochAKNNLS14}. The simplistic query \textbf{countone} is
%intended to explore the minimum intrinsic variation of the system.
%
%The remaining queries are syntactically identical with the experiments
%from~\cite{DBLP:journals/vldb/KochAKNNLS14}.

\begin{small}
\vspace*{-1em}\paragraph{Acknowledgements.}
\label{sec:acknowledgements}
\looseness=-1 
This work was supported by the iDev40 project and the German Research Council (DFG) under grant no. LO 1719/3-1.
The information and results set out in this publication are those of the authors and do not necessarily reflect the opinion of the ECSEL Joint Undertaking.
The iDev40 project has received funding from the ECSEL Joint Undertaking (JU) under grant no.~783163.
The JU receives support from the European Union's Horizon~2020 research and innovation programme.
It is co-funded by the consortium members, grants from Austria, Germany, Belgium, Italy, Spain and Romania.
%
% Machen wir zur CR-Version rein
We thank the DBToaster team, and Jan Kiszka for %technical 
guidance on difficile
technical issues related to Jailhouse on x86 systems.
\end{small}

%%% Angabe der .bib-Datei (ohne Endung) / State .bib file (for BibTeX usage)
\clearpage
\let\section\savesection
%\bibliography{bibliography}
\printbibliography
\end{document}